\documentclass[12pt]{article}

\usepackage{amsmath,amssymb}
\usepackage{cite}

\newcommand{\dd}{\partial}
\newcommand{\de}{\delta}
\newcommand{\m}{\mu}
\newcommand{\n}{\nu}
\newcommand{\ls}{\left(}
\newcommand{\rs}{\right)}
\newcommand{\al}{\alpha}
\newcommand{\be}{\beta}
\newcommand{\ff}{\varphi}

\newcommand{\La}{\Lambda}
\newcommand{\pri}{{\,\in\,}}
\newcommand{\te}{\theta}
\newcommand{\ga}{\gamma}

\newcommand{\disn}[2]{$$\displaylines{\refstepcounter{equation}%
            \label{#1}\hskip 1em minus 1em #2\hfilneg}$$}
\newcommand{\nom}{\hfil\hskip 1em minus 1em (\theequation)}
\newcommand{\no}{\hfil \hskip 1em minus 1em\phantom{(\theequation)}%
            \hfilneg\cr\hfilneg\hskip 1em minus 1em\hfil}

\begin{document}

\title{Embeddings for solutions of Einstein equations}
\author{S.A.~Paston\thanks{E-mail: paston@pobox.spbu.ru},
A.A.~Sheykin\thanks{E-mail: anton.shejkin@gmail.com}\\
{\it Saint Petersburg State University, St.-Petersburg, Russia}
}
\date{\vskip 15mm}
\maketitle
\begin{abstract}
We study isometric embeddings of some solutions of the Einstein equations with sufficiently high symmetries into a flat ambient space. We briefly describe a method for constructing surfaces with a given
symmetry. We discuss all minimal embeddings of the Schwarzschild metric obtained using this method
and show how the method can be used to construct all minimal embeddings for the Friedmann models.
We classify all the embeddings in terms of realizations of symmetries of the corresponding solutions.
\end{abstract}

\newpage
\section{Introduction}
According to the Janet-Cartan theorem \cite{gane,kart}, an arbitrary Riemannian space of dimension $d$ can be
locally embedded isometrically into any Riemann space of dimension $N \geq d(d +1)/2$ and hence into a flat
space of this dimension. Friedman generalized \cite{fridman61} this theorem to the case of a space with a nonpositive
signature. For a four-dimensional space-time, the theorem ensures the existence of a local embedding for
$N = 10$. But in some cases, the ambient space into which we can embed a space-time with a given metric
can have a lower dimension. The number $p=N-d$ for the minimum possible value of $N$ is called the \textit{embedding class} of the metric. The higher the symmetry of a space-time is, the lower its embedding class (see \cite{schmutzer}). On the other hand, if we require that an embedding be global, i.e., that the whole manifold with a given metric be embedded smoothly into the ambient space, then the required dimension $N$ can be much higher than that predicted by the Janet-Cartan theorem because the theorem ensures the existence of only
a \textit{local} embedding. Some results concerning the existence of global embeddings can be found in \cite{kobno}.

If we describe a surface in a flat space using the embedding function $y^a(x^\m)$ (here and hereafter, Greek
indices $\m,\n,\ldots$ take $d$ values and Latin indices  $a,b,\ldots$ take $N$ values), then the problem of constructing
an embedding for the given metric $g_{\m\n}(x)$ reduces to solving the equation
\disn{s2-2}{
\big(\dd_\m y^a\big)\big(\dd_\n y^b\big)\, \eta_{ab}=g_{\m\n},
\nom}
where $\dd_\m\equiv \dd/\dd x^\m$ and $\eta_{ab}$ is the flat metric of an ambient space.

The problem of constructing embeddings for various solutions of the Einstein equations has long been
discussed, almost since general relativity appeared. The purposes of this construction can differ. An explicit
form of the embedding can be useful for better understanding the space-time geometry, which is clearly seen
in the example of Fronsdal's embedding \cite{frons} for the Schwarzschild metric described in Sec.~\ref{shvar} (this embedding
is closely related to the use of Kruskal-Szekeres coordinates; see a remark at the end of \cite{frons}). The existence of
an embedding can be used to solve problems not directly related to embeddings, for example, to find exact solutions of the Einstein equations (see \cite{schmutzer}). The explicit form of an embedding in a physically meaningful
case can be useful when developing the approach in \cite{regge,deser} in which gravity is described not by a metric
but by the embedding function. Such a development relates to studying the possibilities of physically
interpreting extra solutions of the theory \cite{davids97,davids01,statja26}, to studying a canonical structure of the theory\cite{tapia,rojas06,frtap,davkar},
to obtaining a canonical formulation of the theory with additional constraints that ensure the absence of
extra solutions \cite{tmf07-st18,ijtp10-st24}, and to passing to a field theory in the flat ambient space \cite{tmf11-st25}. Some of these results together with a formalism convenient for calculating in the embedding theory framework were presented
in \cite{mon1}. An extensive list of references relevant to embedding theory and related questions can be found
in \cite{tapiaob}.

Constructing explicit embeddings of solutions of the Einstein equations into flat ambient spaces was
discussed in numerous papers devoted mainly to the physically meaningful case $d=4$. An important result
is Kasner's theorem, which states that in this case, a vacuum solution of the Einstein equations (the one
corresponding to a zero Einstein tensor) cannot be embedded into a five-dimensional ambient space, i.e., it
has the embedding class greater than one \cite{kasner2}. Numerous explicit embeddings were collected in \cite{rosen65,collinson68},
and many problems of embeddings constructing were discussed in \cite{goenner}.

Among other embeddings, the problem of embedding the Schwarzschild solution was considered. Kasner's theorem then states that $N\ge 6$ in this case. Until recently, four minimal (i.e., corresponding to
$N = 6$) embeddings of the Schwarzschild metric were known. In \cite{statja27}, we proposed a method for constructing surfaces with a given symmetry and used that method to find all four previously known and two new
minimal embeddings of the Schwarzschild metric. In Sec.~\ref{met}, we briefly describe this method. In Sec.~\ref{shvar},
we present all six minimal embeddings of the Schwarzschild metric obtained using this method, which
have symmetries intrinsic for this metric. In Sec.~\ref{frid}, we describe the results of applying this method to the
problem of constructing embeddings of the Friedmann solution for all three model types: closed, open, and
spatially flat. We reveal the connection between ways of realizing the symmetry for these embeddings and
for the embeddings of the Schwarzschild metric.

\section{The method for constructing surfaces\\ with a given symmetry}\label{met}
We say that a surface $\mathcal M$ is symmetric with respect to a group $G$ if $\mathcal M$ transforms into itself under the
action of a subgroup of $\mathcal P$ isomorphic to $G$, where $\mathcal P$ is the group of motions of the flat space $R^{n_+,n_-}$ (of the flat space with $n_+$ timelike and $n_-$ spacelike directions, $n_++n_-=N>d$). The group of motions of the
space $R^{n_+,n_-}$ is the corresponding generalization of the Poincar$\acute{\text{e}}$ group, i.e., $\mathcal P$ is the semidirect product of
the group $SO(n_+,n_-)$ and the group $T^{n_++n_-}$ of translations of the space $R^{n_+,n_-}$. This definition of the
surface symmetry implies that domains transformed into one another by a symmetry transformation must
have the same internal and external symmetries and can therefore be transformed into one another by a
shift and/or rotation of the ambient space.

To construct the desired surface, we must therefore find a homomorphism $V$ from the group $G$ to the
group $\mathcal P$. The elements of $\mathcal P$ admit a matrix representation in the form of $N+1$ block matrices of the form
 \disn{s2a-0.1}{
\ls
\begin{array}{cc}
\La & a\\
0   & 1\\
\end{array}
\rs,
\nom}
where $\La\pri SO(n_+,n_-)$ and $a\pri R^{n_+,n_-}$ parameterizes the translations (the points of $R^{n_+,n_-}$  are then
described by $(n_++n_-+1)$-vectors whose last component is always one). We can therefore consider $V$ to
be a representation of the group $G$ whose matrices have form (\ref{s2a-0.1}), and the representation is single-valued
and exact but reducible (because matrix (\ref{s2a-0.1}) corresponds to a reducible albeit not completely reducible
representation).
The set of point $y\pri R^{n_+,n_-}$, given by the formula
 \disn{s2a-1}{
y=V(g)y_0
\nom}
for arbitrary $g\pri G$ and for a fixed initial vector  $y_0\pri R^{n_+,n_-}$ is then by construction a surface with the
desired symmetry. If we assume that the vector $y_0$ depends on some continuous parameters, then the set of
points determined by formula (\ref{s2a-1}) is again a surface with this symmetry for all values of these parameters,
and its sections corresponding to fixed parameter values also have this symmetry.

Because the classification of group representations is very well developed, we propose a constructive
way to build surfaces with a given symmetry: we take real-valued representations $V$ of the group $G$ one
after another starting from minimum values of their dimensions and selecting those representations with
representation matrices of form (\ref{s2a-0.1}). We must then take into account that for different representations and
for different initial vectors $y_0$, the obtained surface may have different dimensions. Hence, not all variants
are suitable if the surface dimension $d$ is fixed. We note that the symmetry condition for a desired surface
can be formulated in either global or local form depending on the specific problem.

We can use the presented technique to seek embeddings of Riemannian spaces with a given metric if
their symmetry groups are sufficiently large.
For a fixed $N$, the idea is to use this method to find all surfaces with the symmetry of the metric
and then to impose conditions on the metric of these surfaces. The higher the symmetry
is, the simpler the differential equations are that arise in the last step. For example, when embedding both
the Schwarzschild and the Friedmann metrics, these equations become ODEs (with respect to the radial
parameter $r$ for the Schwarzschild metric and with respect to the time $t$ for the Friedmann metric), and the
problem can be easily solved. If a metric symmetry is not high, then we obtain PDEs, not ODEs, and the
problem becomes much more complicated.

\section{Embeddings of the Schwarzschild metric}\label{shvar}
We now present the results of applying the above technique to the problem of constructing all embeddings of the Schwarzschild metric into the six-dimensional ambient space that has the symmetry of this
metric. We must note that a surface that is an embedding of a space-time might not, in principle, have all
the symmetries intrinsic for this space-time. Such embeddings cannot be found using our technique.

The Schwarzschild metric is a spherically symmetric solution of the Einstein equations in the absence
of matter. The interval of this metric is
 \disn{s2a-2}{
ds^2=\ls 1-\frac{R}{r}\rs dt^2-\frac{dr^2}{1-\dfrac{R}{r}}-r^2\ls d\te^2+\sin^2\te\, d\ff^2\rs,
\nom}
where $R$ is the Schwarzschild radius. The Schwarzschild metric is also invariant under shifts of the coordinate
$t$, and the total symmetry group of a manifold with the Schwarzschild metric is hence $G=SO(3)\times T^1$.

Using the method described in Sec.~\ref{met}, we seek representations of the group $G$ that have matrices written
in form (\ref{s2a-0.1}). Representations of the direct product $SO(3)\times T^1$ can generally be written as the direct sum
of representations that are tensor products of representations of the groups $SO(3)$ and $T^1$. For each of the
obtained representations, in accordance with formula (\ref{s2a-1}), we must write the general form of the embedding
function $y^a(x^\m)$ corresponding to this representation and substitute it in formula (\ref{s2-2}). Solving this equation
thus reduces to solving simple ODEs, which gives explicit possible variants of the embedding function.

It turns out (see the details in \cite{statja27}) that for $N = 6$, a representation $V$ is always the direct sum of
representations of the group $SO(3)$ and $T^1$. The representation of $SO(3)$ can then be only the fundamental
representation, i.e., $SO(3)$ rotations must be realized as the same rotations of a three-dimensional subspace
of the ambient space, while the representation of $T^1$ can be realized in six different ways, which results in
six different embeddings of the Schwarzschild metric for $N=6$.

\textbf{Variant 1}. We realize $t$-shifts as $SO(2)$ rotations in a two-dimensional subspace of the ambient space that as a result have the signature $(+ +)$. The embedding function is
 \disn{s2a-11}{
\begin{array}{lcl}
\displaystyle y^0=\frac{1}{\al}\sqrt{1-\frac{R}{r}}\,\sin(\al t),          &\qquad &  y^3=r\,\cos\te,  \\[1em]
\displaystyle y^1=\frac{1}{\al}\sqrt{1-\frac{R}{r}}\,\cos(\al t),          &\qquad &  y^4=r\,\sin\te\,\cos\ff,\\[1em]
\displaystyle y^2=\int\! dr \sqrt{\frac{R(R+4\al^2 r^3)}{4\al^2 r^3(r-R)}},          &\qquad &  y^5=r\,\sin\te\,\sin\ff\\
\end{array}
\nom}
with the signature  $(++----)$; here, $\alpha$ is an arbitrary positive constant. This embedding was found
by Kasner in 1921 in \cite{kasner3} (he set $\alpha=1$ there), and it is historically the first embedding of the Schwarzschild
metric. It covers only the domain $r>R$ and has a conical singularity at $r = R$. This embedding is not
asymptotically flat, i.e., it does not tend to a four-dimensional plane as $r\to\infty$, and it cannot be written
in terms of elementary functions.

\textbf{Variant 2}. We realize $t$-shifts as $SO(1, 1)$ rotations in a two-dimensional subspace of the ambient
space that as a result have the signature $(+ -)$. The embedding function is
\disn{s2-sh16}{
\begin{array}{lcl}
\qquad\qquad \,\, r>R:        &\qquad & \qquad\qquad \quad  r<R:\\[0.5em]
\displaystyle y^0=2R\,\sqrt{1-\frac{R}{r}}\;\;\sinh\!\ls\frac{t}{2R}\rs, &\qquad &
\displaystyle y^0=\pm 2R\,\sqrt{\frac{R}{r}-1}\;\;\cosh\!\ls\frac{t}{2R}\rs,\\[1em]
\displaystyle y^1=\pm 2R\,\sqrt{1-\frac{R}{r}}\;\;\cosh\!\ls\frac{t}{2R}\rs, &\qquad &
\displaystyle y^1=2R\,\sqrt{\frac{R}{r}-1}\;\;\sinh\!\ls\frac{t}{2R}\rs,\\[1em]
\end{array}\no
\begin{array}{l}
\displaystyle y^2=\int dr \sqrt{\frac{R}{r}+\ls\frac{R}{r}\rs^2+\ls\frac{R}{r}\rs^3},\\[1em]
y^3=r\cos\te,\\
y^4=r\sin\te\,\cos\ff,\\
y^5=r\sin\te\,\sin\ff\\
\end{array}
\nom}
with the signature $(+-----)$. Fronsdal proposed this embedding \cite{frons}. The surface thus defined is
smooth (see \cite{frons} or \cite{statja27}), although this is nonobvious at $r = R$ because the embedding is written in the
Schwarzschild coordinates, which are singular at this point. The Fronsdal embedding differs from other
embeddings because it covers all domains of the Riemannian space corresponding to the Schwarzschild
solution: two copies of the domain $r>R$ and two copies of the domain $r<R$ pertaining to the black
and white holes. It is closely related to the famous Kruskal coordinates (see \cite{statja27}). This embedding is not
asymptotically flat and cannot be written in terms of elementary functions.

\textbf{Variant 3}. We realize $t$-shifts as rotations in a two-dimensional "lightlike" subspace of the ambient
space (in which one direction is timelike and the other is spacelike). The embedding function is
 \disn{s2a-35}{
\begin{array}{lcl}
\displaystyle y^0=t\sqrt{1-\frac{R}{r}},          &\qquad &  y^3=r\,\cos\te,  \\[1em]
\displaystyle y^1=\frac{1}{\sqrt{2}\,\ga}\ls\frac{\ga^2t^2}{2}-1\rs \sqrt{1-\frac{R}{r}}+\frac{u(r)}{\sqrt{2}}, &\qquad &  y^4=r\,\sin\te\,\cos\ff,\\[1em]
\displaystyle y^2=\frac{1}{\sqrt{2}\,\ga}\ls\frac{\ga^2t^2}{2}+1\rs \sqrt{1-\frac{R}{r}}+\frac{u(r)}{\sqrt{2}}, &\qquad &  y^5=r\,\sin\te\,\sin\ff,\\
\end{array}
\nom}
with the signature $(++----)$. Here $\gamma$ is an arbitrary positive constant, and
 \disn{36}{
u(r)=\frac{\ga r(2r+3R)}{4}\sqrt{1-\frac{R}{r}}
+\frac{3\ga R^2}{8}\ln\ls\frac{2r}{R}\ls 1+\sqrt{1-\frac{R}{r}}\rs-1\rs
\nom}
is a function that is smooth for $r>R$. This embedding was first proposed by Fujitani, Ikeda,
and Matsumoto \cite{fudjitani} (there $\ga=\sqrt{2}$). The manifold defined by this embedding covers only the domain $r>R$ and has a conical singularity at $r = R$. This embedding is not asymptotically flat and can be written
in terms of elementary functions.

Depending on the rotation type in the above three variants, we can classify them into the corresponding
elliptic, hyperbolic, and parabolic classes.

\textbf{Variant 4}. We realize $t$-shifts as $SO(2)$ rotations in a two-dimensional subspace of the ambient space accompanied by translations in the direction orthogonal to this subspace. The embedding function is
 \disn{s2a-31}{
\begin{array}{lcl}
\displaystyle y^0=t',          &\qquad &  y^3=r\,\cos\te,  \\[0.5em]
\displaystyle y^1=\frac{(3R)^{3/2}}{\sqrt{r}\quad}\,\sin\ls\frac{t'}{3^{3/2}R}-\sqrt{\frac{R}{r}}\,\bigg( 1+\frac{r}{3R}\bigg)^{3/2}\,\rs, &\qquad &  y^4=r\,\sin\te\,\cos\ff,\\[1.5em]
\displaystyle y^2=\frac{(3R)^{3/2}}{\sqrt{r}\quad}\,\cos\ls\frac{t'}{3^{3/2}R}-\sqrt{\frac{R}{r}}
\,\bigg( 1+\frac{r}{3R}\bigg)^{3/2}\,\rs, &\qquad &  y^5=r\,\sin\te\,\sin\ff\\
\end{array}
\nom}
with the signature $(+-----)$. The parameter $t'=t+h(r)$, where
 \disn{s2a-31a}{
h(r)=3R\sqrt{1+\frac{3R}{r}}+R\,\ln\ls\frac{R\,|r-R|}{\ls r+R+\sqrt{r(r+3R)}\rs^2}\rs,
\nom}
differs from the Schwarzschild time $t$, and the surface determined by the embedding is explicitly smooth. The radiuses of spirals defined by formula (8) at fixed $r$ increase indefinitely as $r \to 0$, i.e., the central singularity moves to infinity. This embedding covers half of the Riemannian space of the Schwarzschild solution determined by the condition $v+u>0$ in the Kruskal-Szekeres coordinates (it covers one copy of the domain $r>R$ and one copy of the domain $r<R$ corresponding to the black hole), and we have $y_0 \to -\infty$ as $v + u \to 0$. In contrast to all other embeddings in the six-dimensional space, this
embedding is asymptotically flat. We proposed this embedding in \cite{statja27} and called it a spiral or asymptotically flat embedding.

\textbf{Variant 5}.  We realize $t$-shifts as $SO(1, 1)$ rotations in a two-dimensional subspace of the ambient space accompanied by translations in the direction orthogonal to this subspace. The embedding function is
\disn{s2a-23}{
\begin{array}{lcl}
\displaystyle y^0=\frac{R}{2\be\sqrt{r_c r}}\ls e^{\be t'+u(r)}-\frac{r-r_c}{R}\,e^{-\be t'-u(r)}\rs,          &\qquad &  y^3=r\,\cos\te,  \\[1em]
\displaystyle y^1=\frac{R}{2\be\sqrt{r_c r}}\ls e^{\be t'+u(r)}+\frac{r-r_c}{R}\,e^{-\be t'-u(r)}\rs, &\qquad &  y^4=r\,\sin\te\,\cos\ff,\\[1em]
\displaystyle y^2=\hat\ga t',            &\qquad &  y^5=r\,\sin\te\,\sin\ff\\
\end{array}
\nom}
with the signature $(+-----)$. Here $\hat{\gamma}>0$, $r_c=R/(1+\hat{\gamma}^2)$, $\beta\ge\sqrt{2r_c/(27R^2(3R-2r_c))}$, $t'=t+h(r)$,
\disn{s2a-23a}{
h(r)=\int\! dr\,\, \frac{R}{2\be r_c(r-R)}\sqrt{\frac{4\be^2r_c^2(r-r_c)}{R-r_c}+\frac{r_c^3(R-r)}{r^3(R-r_c)}},\no
u(r)=\int\! dr\,\, \frac{1-\sqrt{\frac{4\be^2r_c^2(r-r_c)}{R-r_c}+\frac{r_c^3(R-r)}{r^3(R-r_c)}}}{2(r-r_c)}
\nom}
and $u(r)$ is a function smooth at $r>0$. This embedding was first proposed by Davidson and Paz
in \cite{davidson} and was written in form (\ref{s2a-23}) in \cite{statja27}.
The surface determined by this embedding is everywhere smooth ($r\to 0$ as $y^0\to\infty$).

This embedding, just as the spiral embedding, covers half of the Riemannian space corresponding to the Schwarzschild solution, and we have $y^2 \to\infty$ as $v + u \to 0$. It is not asymptotically flat and cannot be written in terms of elementary functions.

\textbf{Variant 6}. We realize $t$-shifts as rotations in a two-dimensional "lightlike" subspace of the ambient space accompanied by a complicated translation. The embedding function is
 \disn{s2a-41}{
\begin{array}{lcl}
\displaystyle y^0=\frac{\xi^2}{6}t'^3+\ls 1-\frac{R}{2r}\rs t'+u(r),     &\qquad &  y^3=r\,\cos\te,  \\[1em]
\displaystyle y^1=\frac{\xi^2}{6}t'^3-\frac{R}{2r}t'+u(r),         &\qquad &  y^4=r\,\sin\te\,\cos\ff,\\[1em]
\displaystyle y^2=\frac{\xi}{2}t'^2+\frac{1}{2\xi}\ls 1-\frac{R}{r}\rs,   &\qquad &  y^5=r\,\sin\te\,\sin\ff\\
\end{array}
\nom}
with the signature $(+-----)$. Here $\xi\ge\sqrt{27}/(32R)$, $t'=t+h(r)$,
 \disn{s2a-41a}{
h(r)=\frac{1}{2\xi}\int dr \frac{\sqrt{Rr\ls \frac{R(R-r)}{r^4}+4\xi^2\rs}}{r-R},\qquad
u(r)=-\frac{1}{2\xi}\int dr \sqrt{\frac{R}{r}\ls \frac{R(R-r)}{r^4}+4\xi^2\rs}
\nom}
and $u(r)$ is a function smooth for $r>0$.
Note that if $\xi=\sqrt{27}/(32R)$ then the integrals (\ref{s2a-41a}) can be expressed through an elliptic functions.
The surface determined by this embedding is everywhere smooth (as $r\to 0$, either $y^0\to\infty$ or $y^1\to\infty$). Just as
the two preceding embeddings, this embedding covers half of the Riemannian space corresponding to the Schwarzschild solution, and we have $y^0\to\infty$ as $v + u \to 0$. It is not asymptotically flat and cannot be written in terms of elementary functions. This embedding was proposed in \cite{statja27}, and we call it the cubic embedding.

We note that because the metric of the Reissner-Nordstr$\ddot{\text{o}}$m charged black hole has the same symmetry as the Schwarzschild metric, we can construct minimal embeddings of the former analogous to those described here including the smooth ones for all $r>0$. Describing these results goes beyond the scope of this paper.

\section{Embeddings of the Friedmann metric}\label{frid}
Because the Friedmann solutions are not vacuum solutions, the Kasner theorem mentioned in Sec.~1
does not restrict their embedding classes, and we have embeddings of these solutions into a five-dimensional
ambient space. Such embeddings for all three Friedmann models (closed, open, and spatially flat) were
presented in \cite{rosen65} but were first found in \cite{robertson}. Some peculiarities of these embeddings were recently discussed
in \cite{smolyakov}.

In this section, we show that these and only these embeddings can be obtained using the above method,
which implies that these embeddings have the symmetry of the corresponding model. While this is clear
for the open and closed models (see formulas (\ref{s2-f2}) and (\ref{s2-f6}) below), the presence of the symmetry group
of the flat three-dimensional space in the four-dimensional surface (determined by formula (\ref{s2-n9}) below)
corresponding to this solution is absolutely nonobvious for the spatially flat model.

\textbf{The closed Friedmann model}.
The metric is
 \disn{s2-f1}{
 ds^2=dt^2-a^2(t)\ls d\chi^2+\sin^2\chi\ls d\te^2+\sin^2\te\, d\ff^2\rs\rs,
\nom}
and the symmetry group is $SO(4)$. We seek a four-dimensional surface $\mathcal M$ with the symmetry with respect
to this group. We then assume that not only the whole surface $\mathcal M$ but also all its three-dimensional
submanifolds with a fixed $t$ parameter have this symmetry.

To construct an embedding into a five-dimensional space, we must find representations $V$ of the group
$SO(4)$ that are either five-dimensional (pseudo)orthogonal matrices or six-dimensional matrices of form (\ref{s2a-0.1}).
Because $V$ must have no continuous kernel, the only admissible representation in this case is the five-dimensional one, which is the direct sum of the vector and scalar representations. The dimension of the obtained surface $\mathcal M$ is then automatically four. Taking into account that the initial vector $y_0$ can depend on $t$ (see the discussion after (\ref{s2a-1})), we eventually obtain the embedding function of the form
\disn{s2-f2}{
\begin{array}{lcl}
y^0=f(t),                 & \qquad  &y^2=a(t)\sin\chi\,\cos\te,\\
y^1=a(t)\cos\chi,         & \qquad  &y^3=a(t)\sin\chi\,\sin\te\,\cos\ff,\\
                          & \qquad  &y^4=a(t)\sin\chi\,\sin\te\,\sin\ff,\\
\end{array}
\nom}
It describes the surface $\mathcal M$ whose submanifolds with $t = \text{const}$ are spheres.
Substituting (\ref{s2-f2}) in (\ref{s2-2}) with the right-hand side corresponding to metric (\ref{s2-f1}), we find that the signature
must be $(+----)$ and that
 \disn{s2-f4}{
f(t)=\int\! dt\, \sqrt{\dot a^2(t)+1},
\nom}
where we let the dot denote the time derivative. Formulas (\ref{s2-f2}) and (\ref{s2-f4}) hence describe the embedding
function for the closed-type Friedmann model solution, and they coincide with those given in \cite{rosen65}.

\textbf{The open Friedmann model}.
The metric is
 \disn{s2-f5}{
 ds^2=dt^2-a^2(t)\ls d\chi^2+\sinh^2\chi\ls d\te^2+\sin^2\te\, d\ff^2\rs\rs,
\nom}
and the symmetry is described by the group $SO(1, 3)$.

We proceed as in the case of the closed model. We must find representations $V$ of the group $SO(1, 3)$
that are either five-dimensional (pseudo)orthogonal matrices or six-dimensional matrices of form (\ref{s2a-0.1}). Again
taking into account that the representation $V$ must have no continuous kernel, we find that the only
suitable representation is again the five-dimensional one, which is the direct sum of the vector and scalar
representations. Because the three-dimensional submanifolds of the desired surface $\mathcal M$ corresponding to
fixed values of $t$ must be spacelike in all three directions, the corresponding embedding function must be
written in the form
\disn{s2-f6}{
\begin{array}{lcl}
y^0=a(t)\cosh\chi,             & \qquad  &y^2=a(t)\sinh\chi\,\sin\te\,\cos\ff,\\
y^1=a(t)\sinh\chi\,\cos\te,    & \qquad  &y^3=a(t)\sinh\chi\,\sin\te\,\sin\ff,\\
                             & \qquad  &y^4=f(t).\\
\end{array}
\nom}
This function describes a surface whose $t=\text{const}$ submanifolds are three-dimensional pseudospheres (hyperboloids).

Substituting (\ref{s2-f6}) in (\ref{s2-2}) with the right-hand side corresponding to metric (\ref{s2-f5}), we find that the signature
must be $(+----)$ and that

 \disn{s2-f8}{
f(t)=\int\! dt\, \sqrt{\dot a^2(t)-1}.
\nom}
We note that the radicand in this formula is always nonnegative because of one of the Friedmann equations. Formulas (\ref{s2-f6}) with (\ref{s2-f8}) taken into account hence provide the embedding function for the open-type
Friedmann solution; these formulas coincide with those given in \cite{rosen65}.

\textbf{The spatially flat Friedmann model}.
The metric is
 \disn{s2-f9}{
 ds^2=dt^2-a^2(t)\ls dr^2+r^2\ls d\te^2+\sin^2\te\, d\ff^2\rs\rs,
\nom}
and the symmetry in this, the most difficult case is the group of motions of the three-dimensional plane
$G=SO(3)\triangleright T^3$, which is the semidirect product of the rotation group $SO(3)$ and the translation group
$T^3$. We can write elements of $G$ in the form $g=\ls O\times b\rs$, where $O\pri SO(3)$ and $b\pri T^3$, i.~e. $b_i$ is the three-dimensional vector parameterizing translations. We seek a four-dimensional surface $\mathcal M$ that is symmetric
with respect to this group and such that its three-dimensional submanifolds corresponding to fixed values
of the parameter $t$ have the same symmetry. We mention that we have other variants in addition to the
expected variant in which these three-dimensional submanifolds are three-dimensional planes (see below).

To construct an embedding into a five-dimensional space, we must find representations $V$ of the group
$G$ that are either five-dimensional (pseudo)orthogonal matrices or six-dimensional matrices of form (\ref{s2a-0.1}). To
describe representations of $G$, we first describe a representation of its subgroup $SO(3)$ and subsequently
define the action of translations $T^3$ on this representation space. As above, we assume that the representation $V$ has no continuous kernel. We can then show that the only possible variant is when $V$ is a
five-dimensional pseudoorthogonal matrix (in some basis). The representation space is then the direct sum of the vector representation and two scalar representations of $SO(3)$, and formula (\ref{s2a-1}) becomes
 \disn{s2-n1}{
y=V(g)\,y_0=
\ls
\begin{array}{c|cc}
O_{ik} & 0 & \be b_i \\ \hline
\be b_m O_{mk} & 1 & \frac{1}{2}\be^2 b_m b_m\\
0 & 0 & 1\\
\end{array}
\rs
\ls
\begin{array}{c}
x^0_k\\ \hline
s^0_1\\
s^0_2\\
\end{array}
\rs,
\nom}
where we write the five-dimensional matrix $V(g)$ in the block form and $\beta$ is an arbitrary constant such that $\beta \neq 0$. The quadratic form preserved by this matrix is up to a factor
  \disn{s2-n2}{
\eta=\ls
\begin{array}{c|cc}
-\de_{ik} & 0 & 0 \\ \hline
0 & 0 & 1\\
0 & 1 & 0\\
\end{array}
\rs,
\nom}
i.e., $V(g)$ in this case is a pseudoorthogonal matrix written in lightlike coordinates.

We can also show that we have a unique variant in which $V$ is a six-dimensional matrix of form (\ref{s2a-0.1}). The
representation space is then the direct sum of one vector representation and three scalar representations of
the $SO(3)$ subgroup, and formula (\ref{s2a-1}) becomes
 \disn{s2-n3}{
y=V(g)\,y_0=
\ls
\begin{array}{c|ccc}
O_{ik} & 0 & 0 & \be b_i \\ \hline
0 & 1 & 0 & 0\\
0 & 0 & 1 & 0\\
0 & 0 & 0 & 1\\
\end{array}
\rs
\ls
\begin{array}{c}
x^0_k\\ \hline
s^0_1\\
s^0_2\\
1\\
\end{array}
\rs.
\nom}
where $\beta \neq 0$. We see that the part of the matrix $V(g)$ corresponding to $\Lambda$ (see formula (\ref{s2a-0.1})) is orthogonal.

We first consider the case of representation (\ref{s2-n3}). Because elements of the group $G$ depend on six real
parameters and the surface corresponding to a fixed value of $t$ must be three-dimensional, we must ensure
that there is a three-dimensional stabilizer group for the initial vector $y_0$. We can easily see that this is the
case only if $x^0_k=0$. The stabilizer group is then $SO(3)$ rotations. As a result, because the initial vector $y_0$
depends on $t$, writing the vector $b_i$ in spherical coordinates $r,\te,\ff$, we find that the embedding function in
this case is
 \disn{s2-n4}{
\begin{array}{lcl}
y^0=s^0_1(t),  & \qquad  &y^2=\be\, r\cos\te,\\
y^1=s^0_2(t),  & \qquad  &y^3=\be\, r\sin\te\,\cos\ff,\\
               & \qquad  &y^4=\be\, r\sin\te\,\sin\ff.\\
\end{array}
\nom}
We stress that $\beta$ here is independent of $t$, and we therefore cannot obtain an embedding of metric (\ref{s2-f9}) using
this embedding function. We note that the section $t = \text{const}$ of the obtained surface is a three-dimensional
plane.

We now consider the case of representation (\ref{s2-n1}). We must again ensure the existence of the three-dimensional stabilizer group of the vector $y_0$, which can be done in two ways. First, we can set  $x^0_k=0$ for $s^0_2\ne 0$, and the stabilizer group is then $SO(3)$ rotations. Second, if $x^0_k\ne 0$, then we necessarily have a one-dimensional stabilizer subgroup $O\pri SO(2)$ corresponding to rotations about the vector $x^0_k$, and we must set $s^0_2=0$ for the total stabilizer group to be three-dimensional. The stabilizer group is then the group of motions of the plane orthogonal to $x^0_k$.

We first consider the variant with $x^0_k\ne 0$. The initial vector $y_0$ must depend on $t$, but the direction of
$x^0_k$, being determined by the stabilizer subgroup of $y_0$, must be independent of $t$. In this case, we therefore
have that only the length of $x^0_k$ (denoted by $f(t)$) and the quantity  $s^0_1(t)$ can depend on t. With $O_{ik}x^0_k$
expressed in terms of $f(t)$ and the spherical angles $\te$ and $\ff$ characterizing $O_{ik}$, it follows from (\ref{s2-n1}) that the
embedding function is
 \disn{s2-n5}{
\begin{array}{lcl}
y^+\!=\be b_m O_{mk}x^0_k+s^0_1(t),  & \qquad  &y^2=f(t)\,\cos\te,\\
y^-\!=0,                             & \qquad  &y^3=f(t)\,\sin\te\,\cos\ff,\\
                                     & \qquad  &y^4=f(t)\,\sin\te\,\sin\ff,\\
\end{array}
\nom}
where  $y^+,y^-$ are lightlike coordinates in the ambient space whose metric is described up to a sign by (\ref{s2-n2}).

We note that for fixed values of the parameters $t,\te$ and $\ff$ , the coordinate $y^+$ can take arbitrary values if we
change the translation parameter $b_m$. We find that embedding function (\ref{s2-n5})) describes a four-dimensional
lightlike plane $y^-=0$, and we therefore cannot obtain an embedding of metric (\ref{s2-f9}) using this embedding
function. We note that the section $t = \text{const}$ of the obtained surface is not a three-dimensional plane,
although we can obtain a three-dimensional plane by taking a different section.

It remains to consider the variant $x^0_k=0$ for representation (\ref{s2-n1}). Again taking the $t$-dependence of
the initial vector $y_0$ into account and writing the vector $b_i$ in the spherical coordinates $r,\te,\ff$, we find that
the embedding function in this case is
 \disn{s2-n6}{
\begin{array}{lcl}
\displaystyle y^+=s^0_1(t)+\frac{1}{2}\be^2s^0_2(t)r^2,  & \qquad  &y^2=\be\, s^0_2(t)\, r\cos\te,\\[1em]
\displaystyle y^-=s^0_2(t),                              & \qquad  &y^3=\be\, s^0_2(t)\, r\sin\te\,\cos\ff,\\[1em]
\displaystyle                                            & \qquad  &y^4=\be\, s^0_2(t)\, r\sin\te\,\sin\ff.\\
\end{array}
\nom}
Substituting it in Eq. (\ref{s2-2}) with the right-hand side corresponding to metric (\ref{s2-f9}), we can find the function
$s^0_{1,2}(t)$ and obtain the final expression (up to the Lorentz boost in the plane $y^+,y^-$) for the embedding
function in the spatially flat Friedmann model:
 \disn{s2-n9}{
\begin{array}{lcl}
\displaystyle y^0=\frac{1}{2}\ls r^2a(t)+\int\frac{dt}{\dot a(t)}+a(t)\rs,  & \qquad  &y^2=a(t)\, r\cos\te,\\[1em]
\displaystyle y^1=\frac{1}{2}\ls r^2a(t)+\int\frac{dt}{\dot a(t)}-a(t)\rs,  & \qquad  &y^3=a(t)\, r\sin\te\,\cos\ff,\\[1em]
\displaystyle                                                               & \qquad  &y^4=a(t)\, r\sin\te\,\sin\ff,\\
\end{array}
\nom}
The signature of the ambient space is $(+----)$. This embedding coincides with that presented
in \cite{rosen65}. We note that a section $t = \text{const}$ (and also all other sections) of the obtained surface is not a three-dimensional plane although it does have a symmetry with respect to the motion group of this plane. This
section is a parabolic surface (because $y^+$ is quadratic in $r$) in the four-dimensional "lightlike" subspace
($y^+,y^2,y^3,y^4$), which is an intermediate variant between the sphere appearing in the closed Friedmann model
and the pseudosphere (hyperboloid) appearing in the open Friedmann model. This is also supported by the
fact that we can obtain embedding (\ref{s2-n9}) by some limit transitions from embeddings (\ref{s2-f2}) and (\ref{s2-f6}) of the
respective closed and open models; this was shown in \cite{statja26} in the example of the closed model.

We can therefore classify the embeddings for the closed, open, and spatially flat Friedmann models as the
respective elliptic, hyperbolic, and parabolic embeddings. With respect to the symmetry realization, these
three embeddings are interrelated as the Kasner, Fronsdal, and Fujitani-Ikeda-Matsumoto embeddings of
the Schwarzschild metric (see variants 1, 2, and 3 in Sec.~\ref{shvar}).

{\bf Acknowledgments}.
The authors thank the organizers of the Fourth International Conference "Models of Quantum Field Theory" (MQFT-2012) dedicated to Alexander Nikolaevich Vasiliev.
This work was supported in part (A.A.Sh.) by the Dynasty Foundation.


\begin{thebibliography}{99}
\bibitem{gane}
M.~Janet, \emph{Ann. Soc. Polon. Math.}, \textbf{5} (1926), 38--43.

\bibitem{kart}
E.~Kartan, \emph{Ann. Soc. Polon. Math.}, \textbf{6} (1927), 1--7.

\bibitem{fridman61}
A.~Friedman, \emph{J. Math. Mech.}, \textbf{10} (1961), 625.

\bibitem{schmutzer}
H.~Stephani, D.~Kramer, M.A.H.~MacCallum, C.~Hoenselaers, and E.~Herlt,
\emph{Exact Solutions of Einstein's Field Equations}, 2nd ed.,
Cambridge Univ. Press, Cambridge, 2003.

\bibitem{kobno}
S.~Kobayashi and K.~Nomizu, \emph{Foundations of Differential Geometry},
Wiley, New York, 1969, v.~2.
\bibitem{frons}
C.~Fronsdal, \emph{Phys. Rev.}, \textbf{116:3} (1959), 778--781.

\bibitem{regge}
T.~Regge, C.~Teitelboim, ``General relativity \`a la string: a progress
report'', \emph{Proceedings of the First Marcel Grossmann Meeting on General Relativity},
(Trieste, Italy, 1975), ed. R.~Ruffini, North-Holland, Amsterdam, 1977, 77--88.

\bibitem{deser}
S.~Deser, F.~A.~E. Pirani, D.~C. Robinson, \emph{Phys. Rev. D}, \textbf{14} (1976), 3301.

\bibitem{davids97}
A.~Davidson, \emph{Class. Quantum. Grav.}, \textbf{16} (1999), 653, arXiv:gr-qc/9710005.

\bibitem{davids01}
A.~Davidson, D.~Karasik, Y.~Lederer, 2001, arXiv:gr-qc/0111107.

\bibitem{statja26}
S.A.~Paston, A.A.~Sheykin, \emph{Int. J. Mod. Phys. D}, \textbf{21:5} (2012), 1250043, arXiv:1106.5212.

\bibitem{tapia}
V.~Tapia, \emph{Clas. Quantum Gravity}, \textbf{6} (1989), L49.

\bibitem{rojas06}
R.~Capovilla, A.~Escalante, J.~Guven, E.~Rojas, \emph{Int. J. Theor. Phys.},
\textbf{48} (2009), 2486, arXiv:gr-qc/0603126.

\bibitem{frtap}
V.A.~Franke, V.~Tapia, \emph{Nuovo Cimento B}, \textbf{107:6} (1992), 611.

\bibitem{davkar}
D.~Karasik, A.~Davidson, \emph{Phys. Rev. D}, \textbf{67} (2003), 064012, arXiv:gr-qc/0207061.

\bibitem{tmf07-st18}
S.~A.~Paston and V.~A.~Franke, \emph{Theor. Math. Phys.}, \textbf{153:2} (2007), 1582--1596, arXiv:0711.0576.

\bibitem{ijtp10-st24}
S.~A.~Paston, A.~N. Semenova, \emph{Int. J. Theor. Phys.}, \textbf{49:11} (2010), 2648--2658, arXiv:1003.0172.

\bibitem{tmf11-st25}
S.A.~Paston, \emph{Theor.~Math.~Phys.}, \textbf{169:2} (2011), 1611--1619, arXiv:1111.1104.

\bibitem{mon1}
S.A.~Paston, \emph{Gravity as Embedding Theory }[in Russian], LAMBERT Academic Publishing, Saarbrucken, 2012.

\bibitem{tapiaob}
M.~Pavsic, V.~Tapia, 2000, arXiv:gr-qc/0010045.

\bibitem{kasner2}
E.~Kasner, \emph{Am. J. Math.}, \textbf{43:2} (1921), 126--129.

\bibitem{rosen65}
J.~Rosen, \emph{Rev. Mod. Phys.}, \textbf{37:1} (1965), 204--214.

\bibitem{collinson68}
C.~D.~Collinson, \emph{J. Math. Phys.}, \textbf{9} (1968), 403.

\bibitem{goenner}
H.~Goenner, "Local isometric embedding of Riemannian manifolds and Einstein's
theory of gravitation", \emph{General Relativity and Gravitation one hundred years after the birth of Albert Einstein}, ed. A.~Held, Plenum Press, 1980, v.~1, ch.~14, pp. 441--468.

\bibitem{statja27}
S.A. Paston, A.A. Sheykin, \emph{Class. Quantum. Grav.}, \textbf{29} (2012), 095022, arXiv:1202.1204.

\bibitem{kasner3}
E.~Kasner, \emph{Am. J. Math.}, \textbf{43:2} (1921), 130--133.

\bibitem{fudjitani}
T.~Fujitani, M.~Ikeda, M.~Matsumoto, \emph{J. Math. Kyoto Univ.}, \textbf{1:1} (1961), 43--61.

\bibitem{davidson}
A.~Davidson, U.~Paz, \emph{Found. Phys.}, \textbf{30:5} (2000), 785--794.

\bibitem{robertson}
H.P.~Robertson, \emph{Rev. Mod. Phys.}, \textbf{5} (1933), 62--90.

\bibitem{smolyakov}
I.~E.~Gulamov, M.N.~Smolyakov, \emph{Gen. Rel. Grav.}, \textbf{44} (2012), 703--710, arXiv:1111.0687.
\end{thebibliography}
\end{document}